\documentclass[final,5p,times,twocolumn,sort&compress]{elsarticle}
\usepackage{amsmath,amssymb,amsfonts,graphicx,slashed,citesort,dsfont}
\usepackage[usenames]{color}
\usepackage{ulem}
\usepackage{nicefrac} 

\newcommand{\be}{\begin{equation}}
\newcommand{\ee}{\end{equation}}
\newcommand{\ba}{\begin{eqnarray}}
\newcommand{\ea}{\end{eqnarray}}
\let\oldhat\hat

\renewcommand{\hat}[1]{\oldhat{\boldsymbol{#1}}}
\allowdisplaybreaks[1]
\begin{document}

\begin{frontmatter}
\title{Electromagnetic Structure of the $Z_c(3900)$}
\author[Bonn]{E.~Wilbring}
\author[Bonn]{H.-W.~Hammer}
\author[Bonn,Julich]{U.-G.~Mei{\ss}ner}

\address[Bonn]{Helmholtz--Institut f\"ur Strahlen- und Kernphysik (Theorie) 
   and Bethe Center for Theoretical Physics, Universit\"at Bonn, D-53115 Bonn, Germany}
\address[Julich]{Institut f\"ur Kernphysik (IKP-3), Institute for Advanced
  Simulation (IAS-4) and J\"ulich Center for Hadron Physics, 
  Forschungszentrum J\"ulich, D-52425  J\"ulich, Germany}

\begin{abstract} 
The observation of the exotic quarkonium state $Z_c(3900)$ by the BESIII
and Belle collaborations supports the concept of hadronic molecules. 
Charmonium states 
interpreted as such molecules would be bound states of heavy particles with 
small binding energies. This motivates their description using an
effective theory with contact interactions. 
In particular, we focus on the electromagnetic
structure of the charged state $Z_c(3900)$. Using first experimental results 
concerning spin and parity, we interpret it as an $S$-wave molecule and 
calculate the form factors as well as charge and magnetic 
radii up to next-to-leading order. We also present first 
numerical estimations of some of these observables at leading order.
\end{abstract}


\end{frontmatter}


\section{Introduction}

In 2013 the BESIII and Belle collaborations reported the observation of a charged state 
in the charmonium sector called $Z_c(3900)$ \cite{Ablikim:2013mio,Liu:2013dau} (this state
was also seen at CLEO \cite{Xiao:2013iha}). The measured mass and width are
\begin{eqnarray}\label{masses_and_widths}
 M_{Z} = (3899.0 \pm 8.5) \: \rm{MeV} & & \Gamma_{Z} = (46 \pm 30) \: \rm{MeV} \:.
\end{eqnarray}
Soon after its discovery it was proposed that these states are $S$-wave
hadronic molecules \cite{Wang:2013cya, Guo:2013sya}. Later a further analysis in this
framework and different approaches for the substructure like tetraquarks 
were presented in Ref.~\cite{Voloshin:2013dpa}.
In this work, we will assume that the $Z_c$ is indeed a hadronic molecule. 
A comparison with combinations of charm mesons shows that this states lies
close to a specific threshold and thus favors this interpretation. Denoting 
the wave function with the respective particle name, one can identify their 
main component to be \cite{Wang:2013cya, Voloshin:2013dpa}
\begin{eqnarray}
 Z_c & = & \frac{1}{\sqrt{2}} \Bigl(\bar{D} D^* + D \bar{D}^* \Bigr) \:.
\end{eqnarray}
First, the constituents are charm mesons whose masses are far above
$1\,\rm{GeV}$ and second, their binding energy is rather small, i.e. 
below the pion mass $m_{\pi}$ or at least of that order. Hence, for the 
description of these states one can use in a first approximation a 
pionless effective field theory  where the constituents are treated as 
non-relativistic, point-like particles which only interact via contact 
interactions. Such an short-range effective theory is called
EFT($\slashed{\pi}$) and is a commonly used tool in nuclear physics 
\cite{van_Kolck:1999mw, Beane:2000fx, Bedaque:2002mn, Epelbaum:2008ga}. Its expansion parameter is 
${Q}/{m_\pi}$, where $Q$ is the scale of the modulus $p$ of the internal 
momenta $\mathbf{p}$ of the involved particles which are - according to the 
small binding energy of the molecule, $|B| \lesssim 30\: \rm{MeV}$ - below the 
pion mass $m_{\pi}$. We already remark here that the characteristics of the
$Z_c$ are such that EFT($\slashed{\pi}$) is barely applicable, but it
still allows
for some first insights into the electromagnetic structure of this exotic
particle. 

In particular, using the charm meson masses from Ref.~\cite{Nakamura:2010zzi} 
the (averaged) binding energy $B$ can be obtained as
\begin{eqnarray}
 B & = & \frac{1}{2} \left[(M_+ + M_{*0} - M_Z) \: + \: (M_0 + M_{*+} - M_Z) \right] \nonumber\\
   & = & (-23.13 \: \pm \: 6.01) \: \rm{MeV} \label{binding_energy_Z_c} \:.
\end{eqnarray}
The negative value implies a resonance instead of
a charm meson bound state solution.
Neglecting the imaginary part of the binding momentum $\gamma$
(cf. Sec.~\ref{sec:forma}), we treat the $Z_c(3900)$ as a 
virtual state with negative scattering length. 
The binding momentum obtained from the extracted binding 
energy, Eq.~(\ref{binding_energy_Z_c}), is
\begin{eqnarray}
\label{binding_momentum}
\gamma &=& (-211.6 \: \pm \: 27.5) \: \rm{MeV}~, 
\end{eqnarray}
where the uncertainty is entirely from the one in the binding 
energy. 
In leading order, the 
electromagnetic properties are independent of the sign of $\gamma$. Our EFT expansion parameter is ${Q}/{m_{\pi}}$ 
and since the binding momentum 
$\gamma$ should be a low-energy scale proportional 
to $Q$ to provide a proper counting scheme, one concludes that 
$|\gamma|$ must be smaller than 
${m_{\pi}} \approx 140~\rm{MeV}$. 
This is not the case for our 
estimate of $\gamma$ in Eq.~(\ref{binding_momentum}). 
However, one can use the predictions for the $Z_c$ 
as first, {\it rough} model estimations and we will do so in the
following. Our predictions 
can be made more quantitative by explicitly including pions
in the theory. 


For the charm molecule $Z_c$ the analysis of its decay channel shows that the 
quantum numbers $I(J^P) = 1(1^+)$ are favored \cite{Ablikim:2013mio, Wang:2013cya}, which 
agrees with the $S$-wave hypothesis. Thus the $Z_c$ can be described by 
an effective field theory with $S$-wave contact interactions similar
to the loosely-bound charm molecule $X(3872)$ 
\cite{Braaten:2003he,AlFiky:2005jd,Canham:2009zq}.
The structure of the $Z_c$ state can then be calculated analog to 
the deuteron treated in EFT($\slashed{\pi}$) as a baryonic molecule of 
nucleons in a relative $S$-wave with total spin $J=1$ 
\cite{Chen:1999tn, Phillips:2002da, Beane:2000fi, Kaplan:1998sz}. Besides this
analogy, there is also a conceptual connection to the so-called Halo EFT
developed in Refs.~\cite{Bertulani:2002sz, Bedaque:2003wa} and in particular 
to its application in Ref.~\cite{Hammer:2011ye}.


\section{Formalism}
\label{sec:forma}
Following Refs.~\cite{Canham:2009zq,Chen:1999tn, Hammer:2011ye}, 
one can write the effective Lagrangian up 
to next-to-leading order (NLO) for the positively charged $Z_c$ as
\begin{eqnarray} \label{Z_Lagrangian}
 \mathcal{L} & = & (D^{+})^{\dagger} \left(i \partial_t + eA_0 + 
\frac{(\nabla+ie\mathbf{A})^2}{2M_{+}} \right) D^{+} \nonumber\\
&& +  (\bar{D}^{0})^{\dagger} \left(i \partial_t + \frac{\nabla^2}{2M_{0}} 
\right) \bar{D}^{0}\nonumber\\
& & + (D^{*+})^{\dagger} \left(i \partial_t + eA_0 + 
\frac{(\nabla+ie\mathbf{A})^2}{2M_{*+}} \right) D^{*+} \nonumber \\ 
&& + (\bar{D}^{*0})^{\dagger} 
\left(i \partial_t + \frac{\nabla^2}{2M_{*0}} \right) \bar{D}^{*0} \nonumber \\
& & + Z^{\dagger} \left[\eta \left(i \partial_t + eA_0 + 
\frac{(\nabla+ie\mathbf{A})^2}{2M_{Z}} \right) + \Delta \right]Z \nonumber\\ 
& & - g \biggl[Z^{\dagger} \Bigl(D^{+} \bar{D}^{*0} + D^{*+}
\bar{D}^{0}\Bigr) \: + \: \Bigl(D^{+} \bar{D}^{*0} 
+ D^{*+} \bar{D}^{0}\Bigr)^{\dagger} Z \biggr] \nonumber\\
& & - \mu_{+} (D^+)^{\dagger} \mathbf{U} \cdot \mathbf{B} D^+ \: 
- \: \mu_{*+} (D^{*+})^{\dagger} \mathbf{U} \cdot \mathbf{B} D^{*+} \nonumber\\ 
&& - \mu_Z Z^{\dagger} \mathbf{U} \cdot \mathbf{B} Z \:.
\end{eqnarray}
This Lagrangian and our formalism are generic for 
non-relativistic particles. The kinetic terms for the $D$ and $D^*$
mesons simply reproduce the free Schr\"odinger equation which is linear 
in the time derivative. 
Moreover, $Z$ is an auxiliary field describing the molecular state $Z_c(3900)$, 
$g$ is the coupling constant for $ZDD^*$ interactions, $\Delta$ is a 
constant and $\eta = \pm 1$ is a phase whose sign depends on the 
effective range. 

We stress that the $Z_c$ is not treated as a dynamical degree of freedom. 
We use the "dimeron" formalism of Kaplan \cite{Kaplan:1996nv}
which is convenient to treat non-perturbative contact interactions.
If the auxilliary field is integrated out, the Langrangian (\ref{Z_Lagrangian})
reduces to one with contact interactions between the $D$ and $D^*$ mesons.
If the phase $\eta$ is negative, the $Z$ field is a ghost but 
the scattering amplitude for the $D$ and $D^*$ mesons is unitary and
sensible. This method is well established in effective field theory
treatments of 
resonant interactions in nuclear and particle physics \cite{Hammer:2010kp}.

Note also that the spin structure of the $D^*$ and $Z_c$ is hidden
in Eq.~(\ref{Z_Lagrangian}) since the interactions are
spin-independent. Since there are only $S$-wave interactions,
the spin state of the $Z_c$ is always equal 
to the spin state of the constituent $D^*$ and remains unchanged in the 
$ZDD^*$ interaction vertex.

Minimal substitution with electric charge $e$ induces 
an $A_0$ interaction with the constituents and the molecule itself, 
leading to the Feynman diagrams which contribute to the charge form factor. 
The magnetic form factor instead comes from the terms in the last 
line of Eq.~(\ref{Z_Lagrangian}), where $\mathbf{B}= \mathbf{\nabla} \times 
\mathbf{A}$ is the magnetic field and $\mathbf{U}$ is the spin-1
generalization of the Pauli vector for spin-1/2 particles, with 
$(U_k)_{jl}=-i\epsilon_{kjl}$ being the generators 
of the rotation group. Those terms are proportional to the magnetic moment 
$\mu$ of the respective charged particle and describe purely magnetic 
interactions of the spatial part of the vector potential $A_i$ with both 
the charged $D$ mesons and $Z_c$. Other contributions to the Lagrangian, 
for instance those relevant for the quadrupole form factor or effects due 
to the anomalous magnetic moment of neutral $D$ mesons, 
are at least one order higher and thus not shown. 
Finally, we note that the $Z_c$ binding energy 
scale of order 20 MeV is small compared to the $D$-$D^*$ mass splitting 
of about 140 MeV.  Our Lagrangian is only valid in a small energy region 
close to the $Z_c$ mass and thus heavy quark symmetry is not relevant.

\begin{figure}[tb]
\begin{center}
    \includegraphics[width=0.475\textwidth]{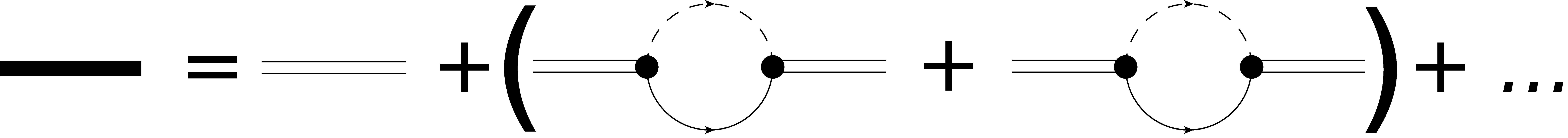}
\end{center}
\caption{\label{Dyson_Zc}Full $Z$ propagator (thick solid line) as sum over
  charm meson loops. The double solid line represents the bare $Z$ propagator, 
  the single solid line stands for $D^{*}$ mesons, the dashed one for $D$ 
  mesons and the dots represent higher order terms.}
\end{figure}

The auxiliary field $Z$ is not dynamical but due to its coupling to the
constituents it is dressed by $D$ meson loops, see Fig.~\ref{Dyson_Zc}.
This leads to a full $Z$ propagator
\begin{eqnarray}
 iS_{Z} = \frac{i}{\left(S_{Z}^0\right)^{-1} - \Sigma} \:,
\end{eqnarray}
in terms of the bare propagator $iS_{Z}^0$ and the self-energy $\Sigma$. 
To avoid an unnatural scaling of the scattering length (similar to 
that discussed in Ref.~\cite{Kaplan:1996xu}), 
one calculates the self-energy in the
PDS scheme introduced in Refs.~\cite{Kaplan:1998tg, Kaplan:1998we};
see Ref.~\cite{vanKolck:1998bw} for an alternative solution to this
problem.
In terms of the 
PDS scale $\Lambda_{\rm PDS}$ and the mean reduced mass
$\omega_{Z}=\frac{1}{2} 
(\frac{M_{+} M_{*0}}{M_{+} + M_{*0}} + \frac{M_{0} M_{*+}}{M_{0} + M_{*+}})$
of the molecule, one obtains
\begin{eqnarray}
 \Sigma = -\frac{\omega_{Z} g^2}{\pi} \left[i \sqrt{2\omega_{Z} \left(p_0 -
       \frac{p^2}{M_{Z}} \right) + i\varepsilon} + \Lambda_{\rm PDS} \right] \:.
\end{eqnarray}
From the full propagator of the $Z$ field at NLO,
\begin{eqnarray} \label{full_propagator_Z}
 iS_{Z}(p_0, \mathbf{p}) & = & \frac{\pi i}{\omega_{Z} g^2} \Biggl[\frac{\pi \Delta}{\omega_{Z} g^2} \: + \: \frac{\pi \eta}{ \omega_{Z} g^2} \left(p_0 - \frac{p^2}{2M_{Z}} \right) \nonumber \\
& + & \Lambda_{\rm PDS}  + \: i \sqrt{2\omega_{Z} \left(p_0 
- \frac{p^2}{2M_{Z}} \right) +i \varepsilon} \: \Biggr]^{-1} \:,
\end{eqnarray}
one can deduce the scattering length $a$ and the effective range $r_0$ of the
$Z_c$ by matching the scattering amplitude $-iT ~=~ (-ig)^2 iS_{Z}(E =
{k^2}/({2\omega_{Z}}), 0)$ 
to the effective range expansion (ERE)
\begin{eqnarray}
 T_{ERE}^{(2)} = \frac{\pi}{\omega_Z} \frac{1}{\frac{1}{a} - \frac{1}{2}r_0k^2 + ik} \:.
\end{eqnarray}
At NLO the following relations are obtained:
\begin{eqnarray}
 \frac{1}{a} & = & \frac{\pi \Delta}{\omega_{Z} g^2} + \Lambda_{PDS}~, \\
 r_0 & = & - \frac{\pi \eta}{\omega_{Z}^2 g^2} \:.
\end{eqnarray}
In addition this result fixes the phase introduced in the Lagrangian 
Eq.~(\ref{Z_Lagrangian}). To get a positive value for the effective range, 
the phase $\eta$ must be equal to minus one.

Furthermore, we calculate the wave function renormalization constant $Z$ 
defined as the residue of the pole of the propagator in 
Eq.~(\ref{full_propagator_Z}). At  NLO one finds
\begin{eqnarray}
 Z_{NLO} & = & \frac{\pi \gamma}{\omega_{Z}^2 g^2} \frac{1}{1-r_0 \gamma} \label{wfrc_NLO} \:,
\end{eqnarray}
where the $Z_c$ binding momentum $\gamma = {\rm sgn(B)}
\sqrt{2\omega_{Z} |B|}$ was
introduced whose definition is chosen such that one takes care 
of both, bound {\it and} virtual states.\\

The non-relativistic matrix elements of the electromagnetic current $J^{\mu}$ 
can be expanded in terms of form factors. For a vector particle state 
$\left|\mathbf{p}, \varepsilon_i \right>$ with three momentum $\mathbf{p}$ 
and linear polarization vector $\varepsilon_i$, one can write in the 
Breit--frame\cite{Chen:1999tn, Hammer:2011ye}
\begin{eqnarray}
 \langle \mathbf{p'},\varepsilon_j\left|J^0\right|\mathbf{p},\varepsilon_i\rangle 
& = & -ie\left[G_c(q) \delta_{ij} \right.\nonumber \\
&&  \left. + \frac{1}{2M^2} G_q(q) \left(q_i q_j - \frac{q^2
      \delta_{ij}}{n-1}\right)\right]~,
\label{J^0-current_Breitframe}\\
 \langle \mathbf{p'},\varepsilon_j\left|J^k\right|\mathbf{p},\varepsilon_i\rangle 
& = & -i\frac{e}{2M} \biggl[G_m(q) (\delta^k_j q_i - \delta^k_i q_j)\biggr] 
\label{J^k-current_Breitframe}\:.
\end{eqnarray}
Here, $\mathbf{q} = \mathbf{p'} - \mathbf{p}$ is the three-momentum transfer 
with $q_{\mu}q^{\mu} = q_0^2 - q^2$ and $n$ is the number of space-time
dimensions (here $n=4$). The charge ($G_c$), magnetic ($G_m$) and quadrupole
($G_q$) form factors are normalized in the following way \cite{Zuilhof:1980ae}:
\begin{eqnarray} \label{formfactor_normalization}
G_c(q=0) &=& 1~,\nonumber\\
\frac{e}{2M} G_m(q=0) &=& \mu~, \nonumber\\
\frac{1}{M^2} G_q(q=0) &=&
 \kappa \:,
\end{eqnarray}
in terms of  the magnetic moment $\mu$ of the particle and its quadrupole
moment $\kappa$.

\begin{figure}[tb]
\begin{center}
    \includegraphics[width=0.48\textwidth]{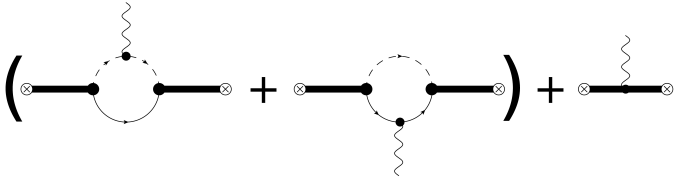}
\end{center}
\caption{\label{G_c_NLO_general_Zc}{Diagrams contributing to the charge and magnetic form factor of the $Z_c$ at NLO. While the left diagrams in parenthesis are LO, the right diagram is the NLO contribution. The thick solid line represents the full $Z$ propagator, the thin solid line stands for $D^{*}$ and the dashed line for $D$ mesons. The incoming photons are indicated by wavy lines and the application of wave function renormalization factors is indicated by the crosses.}}
\end{figure}

According to Eq.~(\ref{J^0-current_Breitframe}), the charge form factor is
determined by the diagrams shown in Fig.~\ref{G_c_NLO_general_Zc} in which the 
incoming photons are $A_0$ photons and where the initial and final spin 
of the $Z_c$ are equal.
The charge form factor at NLO of the $Z_c$ is then given by
\begin{eqnarray}
 G^{(NLO)}_c(q) & = & \frac{1}{1-\gamma r_0} 
\left[\frac{M_{+} \gamma}{q \omega_{Z}} \arctan \left(\frac{q
      \omega_{Z}}{2M_{+} \gamma}\right) \right. \nonumber\\
&+& \left. \frac{M_{*+} \gamma}{q \omega_{Z}} \arctan \left(\frac{q
      \omega_{Z}}{2M_{*+} 
\gamma}\right) \: - \: \gamma r_0 \right] \:. 
\end{eqnarray}
This is indeed correctly normalized to one. A second observable besides the 
form factor itself is the expectation value of the squared charge radius $\langle r_c^2 \rangle$. It can be found 
from the low-energy expansion of the form factor,
$G_c (q) = 1 - \langle r_c^2 \rangle q^2/6 + {\cal O}(q^4)$. At NLO one finds
\begin{eqnarray}
 \langle r^2_c \rangle & = & \frac{1}{1-\gamma r_0} \frac{\omega_{Z}^2}{4\gamma^2} \left(\frac{1}{M_{+}^2} + \frac{1}{M_{*+}^2}\right) \:.
\end{eqnarray}
If the incoming photons in Fig.~\ref{G_c_NLO_general_Zc} are $A_i$ photons, 
one can choose the polarizations in Eq.~(\ref{J^k-current_Breitframe}) and 
calculate the amplitude of the corresponding diagrams to get the magnetic 
form factor 
$G_m$ of the $Z_c$. At NLO it is given by
\begin{eqnarray}
 G_m^{(NLO)}(q) &=& \frac{2M_{Z}}{e(1-\gamma r_0)} \: \left[
   \frac{\mu_{+}}{2} \frac{M_{+}\gamma}{q\omega_{Z}} \arctan
   \left(\frac{q\omega_{Z}}{2M_{+}\gamma}\right) \right. \nonumber\\
&+& \left. \frac{\mu_{*+}}{2} \frac{M_{*+}\gamma}{q\omega_{Z}} 
\arctan \left(\frac{q\omega_{Z}}{2M_{*+}\gamma}\right) \: - \: \gamma r_0
\mu_{Z} \right] \:.\nonumber \\ &&
\end{eqnarray}
So the difference to the charge form factor are the magnetic moments 
in front of each term. With the normalization condition in 
Eq.~(\ref{formfactor_normalization}). it is possible to identify a
relation between the magnetic moment of the molecule $Z_c$ and that of its 
charged constituents $D^+$ and $D^{*+}$
\begin{eqnarray}\label{moments}
 \mu_{Z} = \frac{1}{4} \left(\mu_+ + \mu_{*+}\right) \:.
\end{eqnarray}
In addition, one can determine the expectation value of the squared magnetic 
radius $\langle r_m^2 \rangle$ by expanding the magnetic form factor up to second order in
$q$. At NLO one finds
\begin{eqnarray}
 \langle r_m^2 \rangle & = & \frac{1}{1-\gamma r_0} \frac{\omega_{Z}^2}{4\gamma^2} 
\left[\frac{1}{\mu_+ + \mu_{*+}} \: \left(\frac{\mu_+}{M_+} 
+ \frac{\mu_{*+}}{M_{*+}}\right) \right] \:. 
\end{eqnarray}
According to Eq.~(\ref{J^0-current_Breitframe}), the quadrupole form factor 
is determined by $A_0$ photon interactions in Fig.~\ref{G_c_NLO_general_Zc} 
with a spin change between the initial and the final state. The $A_0$
interaction itself cannot induce such a spin change because $A_0$ is just 
the time-like component of the $4$-vector potential $A_{\mu}$. Thus one needs 
additional operators in the Lagrangian which project on specific polarization 
states $\varepsilon_i$. According to Ref.~\cite{Chen:1999tn} 
such operators first appear at 
NLO and thus the quadrupole form factor $G_q(q)$ vanishes at leading order.


\section{Results}

\label{section_results}
Except for the masses, widths and $I(J^P)$ there are no experimental data available for the 
$Z_c(3900)$, hence one can predict observables only at LO. Furthermore, the leading order 
magnetic form factor and the magnetic radius are proportional to the magnetic moments 
of the constituent mesons. As these are experimentally unknown both are not accessible 
even at LO. Thus only the charge form factor and the charge radius at LO can be predicted.

The results for the scattering length and
charge radius of the $Z_c$ 
obtained from the extracted binding momentum,
Eq.~(\ref{binding_momentum}) are
\begin{eqnarray}
\label{scattering_lengths_and_charge_radii}
a &=& (-0.93 \: \pm \: 0.12) \: \rm{fm}~, \nonumber\\ 
\langle r_c^2\rangle &=& (0.11 \: \pm 0.03)\: \rm{fm^2}~,
\end{eqnarray}
where the uncertainties are entirely from the one in the binding 
energy, Eq.~(\ref{binding_energy_Z_c}). 
Due to the large expansion parameter,
we refrain from estimating the errors due to higher orders.
We reiterate that our predictions for the $Z_c$ should thus be
considered model estimates.
Also, note that the charge radius in Eq.~(\ref{scattering_lengths_and_charge_radii})
should be added in quadrature the charge radii of the constituent $D^{(*)}$ mesons which
are treated as pointlike in our theory.

\begin{figure}[t]
\begin{center}
    \includegraphics[width=0.49\textwidth]{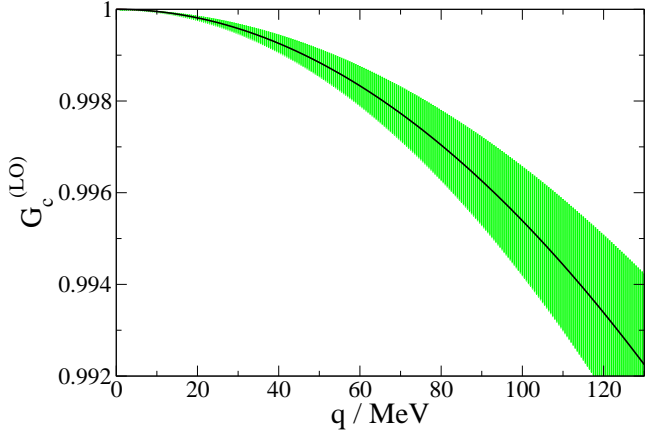}
\end{center}
\caption{\label{Z_c_G_c_LO_plot}{LO charge form factor $G_c^{(LO)}(q)$ of 
                the $Z_c(3900)$. The error band includes only the uncertainty
 in the binding energy and is depicted in green.}}
\end{figure}

\begin{figure}[htb]
\begin{center}
    \includegraphics[width=0.49\textwidth]{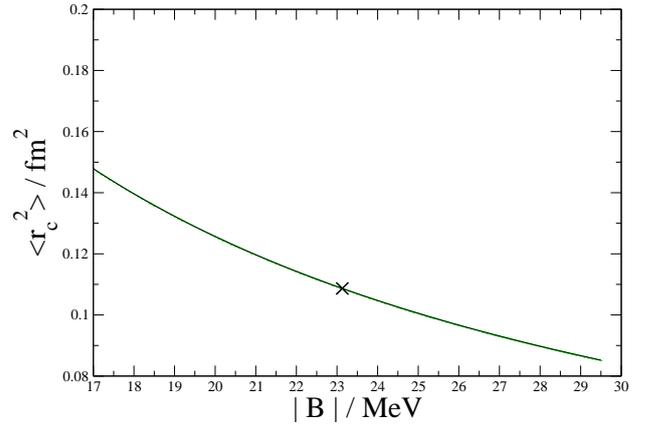}
\end{center}
\caption{\label{Charge-radius_correlation_Z_c_plot}{Correlation between the
    expectation value of the squared charge radius $\langle r_c^2 \rangle$ 
    and the modulus of the binding energy $B$ of the $Z_c(3900)$ at LO. 
    The correlation is shown for binding energies within the errors of 
    Eq.~(\ref{binding_energy_Z_c}). The central value 
    Eq.~(\ref{scattering_lengths_and_charge_radii}) is represented by the cross 
    and the error band is depicted in green.}}
\end{figure}

The charge form factor of the $Z_c$ at LO
is shown in Fig.~\ref{Z_c_G_c_LO_plot}. 
Furthermore, there is a the correlation between the leading order charge 
radius and the binding energy as shown  in 
Fig.~\ref{Charge-radius_correlation_Z_c_plot}.

%

\section{Summary}
In this work the electromagnetic structure of the exotic meson $Z_c(3900)$ 
interpreted as an $S$-wave charm meson molecule with spin $J=1$ was investigated
in the framework of  EFT($\slashed{\pi}$). Similar to the description of the
deuteron in this effective theory, the charge, magnetic and quadrupole form
factor were formally analyzed up to NLO. In particular,  the charge and magnetic radius 
were obtained. Due to the poor experimental data on the $Z_c$ state only the 
LO charge radius and charge form factor could be estimated. Furthermore, we
derived a relation between the magnetic moment of the molecule and of its 
constituents, see Eq.~(\ref{moments}). Given the experimental mass of the 
$Z_c(3900)$, the
binding momentum of the $Z_c(3900)$ is a bit too large to allow for a
quantitative description of its properties in  EFT($\slashed{\pi}$). 
Thus our results should be viewed as as first, {\it rough} model estimations
of the electromagnetic properties of the $Z_c(3900)$. To sharpen these predictions, 
one has to go beyond the pionless EFT and include explicit pions.

\vspace*{0.2cm}

\noindent
\textbf{Acknowledgments}
This work is supported in part by the EU Integrated Infrastructure Initiative
HadronPhysics3 and the 
DFG  through funds provided to the Sino-German CRC~110 ``Symmetries and the
Emergence of Structure in QCD''.



\begin{thebibliography}{99}

\bibitem{Ablikim:2013mio} 
  M.~Ablikim {\it et al.}  [ BESIII Collaboration],
  arXiv:1303.5949 [hep-ex].

\bibitem{Liu:2013dau}
  Z.~Q.~Liu {\it et al.}  [Belle Collaboration],
  arXiv:1304.0121 [hep-ex].


\bibitem{Xiao:2013iha}
  T.~Xiao, S.~Dobbs, A.~Tomaradze and K.~K.~Seth,
  arXiv:1304.3036 [hep-ex].



\bibitem{Wang:2013cya} 
  Q.~Wang, C.~Hanhart and Q.~Zhao,
  arXiv:1303.6355 [hep-ph].


\bibitem{Guo:2013sya} 
  F.~-K.~Guo, C.~Hidalgo-Duque, J.~Nieves and M.~P.~Valderrama,
  arXiv:1303.6608 [hep-ph].


\bibitem{Voloshin:2013dpa} 
  M.~B.~Voloshin,
  arXiv:1304.0380 [hep-ph].


\bibitem{van_Kolck:1999mw} 
  U.~van Kolck,
  Prog.\ Part.\ Nucl.\ Phys.\  {\bf 43}, 337 (1999)
  [nucl-th/9902015].


\bibitem{Beane:2000fx} 
  S.~R.~Beane, P.~F.~Bedaque, W.~C.~Haxton, D.~R.~Phillips and M.~J.~Savage,
  In *Shifman, M. (ed.): At the frontier of particle physics, vol. 1* 133-269
  [nucl-th/0008064].


\bibitem{Bedaque:2002mn} 
  P.~F.~Bedaque and U.~van Kolck,
  Ann.\ Rev.\ Nucl.\ Part.\ Sci.\  {\bf 52}, 339 (2002)
  [nucl-th/0203055].


\bibitem{Epelbaum:2008ga} 
  E.~Epelbaum, H.-W.~Hammer and U.-G.~Mei{\ss}ner,
  Rev.\ Mod.\ Phys.\  {\bf 81}, 1773 (2009)
  [arXiv:0811.1338 [nucl-th]].

\bibitem{Nakamura:2010zzi} 
  J.~Beringer {\it et al.}  [Particle Data Group Collaboration],
  Phys.\ Rev.\ D {\bf 86}, 010001 (2012).


\bibitem{Braaten:2003he} 
  E.~Braaten and M.~Kusunoki,
  Phys.\ Rev.\ D {\bf 69}, 074005 (2004)
  [hep-ph/0311147].

\bibitem{AlFiky:2005jd} 
  M.~T.~AlFiky, F.~Gabbiani and A.~A.~Petrov,
  Phys.\ Lett.\ B {\bf 640}, 238 (2006)
  [hep-ph/0506141].

\bibitem{Canham:2009zq} 
  D.~L.~Canham, H.-W.~Hammer and R.~P.~Springer,
  Phys.\ Rev.\ D {\bf 80}, 014009 (2009)
  [arXiv:0906.1263 [hep-ph]].

\bibitem{Chen:1999tn} 
  J.~-W.~Chen, G.~Rupak and M.~J.~Savage,
  Nucl.\ Phys.\ A {\bf 653}, 386 (1999)
  [nucl-th/9902056].


\bibitem{Phillips:2002da} 
  D.~R.~Phillips,
  Czech.\ J.\ Phys.\  {\bf 52}, B49 (2002)
  [nucl-th/0203040].


\bibitem{Beane:2000fi} 
  S.~R.~Beane and M.~J.~Savage,
  Nucl.\ Phys.\ A {\bf 694}, 511 (2001)
  [nucl-th/0011067].


\bibitem{Kaplan:1998sz} 
  D.~B.~Kaplan, M.~J.~Savage and M.~B.~Wise,
  Phys.\ Rev.\ C {\bf 59}, 617 (1999)
  [nucl-th/9804032].


\bibitem{Bertulani:2002sz} 
  C.~A.~Bertulani, H.-W.~Hammer and U.~Van Kolck,
  Nucl.\ Phys.\ A {\bf 712}, 37 (2002)
  [nucl-th/0205063].


\bibitem{Bedaque:2003wa} 
  P.~F.~Bedaque, H.-W.~Hammer and U.~van Kolck,
  Phys.\ Lett.\ B {\bf 569}, 159 (2003)
  [nucl-th/0304007].

\bibitem{Hammer:2011ye} 
  H.-W.~Hammer and D.~R.~Phillips,
  Nucl.\ Phys.\ A {\bf 865}, 17 (2011)
  [arXiv:1103.1087 [nucl-th]].


\bibitem{Kaplan:1996nv} 
  D.~B.~Kaplan,
  Nucl.\ Phys.\ B {\bf 494}, 471 (1997)
  [nucl-th/9610052].


\bibitem{Hammer:2010kp} 
  H.-W.~Hammer and L.~Platter,
  Ann.\ Rev.\ Nucl.\ Part.\ Sci.\  {\bf 60}, 207 (2010)
  [arXiv:1001.1981 [nucl-th]].

\bibitem{Kaplan:1996xu} 
  D.~B.~Kaplan, M.~J.~Savage and M.~B.~Wise,
  Nucl.\ Phys.\ B {\bf 478}, 629 (1996)
  [nucl-th/9605002].


\bibitem{Kaplan:1998tg} 
  D.~B.~Kaplan, M.~J.~Savage and M.~B.~Wise,
  Phys.\ Lett.\ B {\bf 424}, 390 (1998)
  [nucl-th/9801034].


\bibitem{Kaplan:1998we} 
  D.~B.~Kaplan, M.~J.~Savage and M.~B.~Wise,
  Nucl.\ Phys.\ B {\bf 534}, 329 (1998)
  [nucl-th/9802075].

\bibitem{vanKolck:1998bw} 
  U.~van Kolck,
  Nucl.\ Phys.\ A {\bf 645}, 273 (1999)
  [nucl-th/9808007].


\bibitem{Zuilhof:1980ae} 
  M.~J.~Zuilhof and J.~A.~Tjon,
  Phys.\ Rev.\ C {\bf 22}, 2369 (1980).




\end{thebibliography}
\end{document}